\documentclass[letter]{IEEEtran}
\usepackage{amsmath,amssymb,amsfonts}
\usepackage{algorithmic}
\usepackage{algorithm}
\usepackage{array}
\usepackage[caption=false,font=normalsize,labelfont=sf,textfont=sf]{subfig}
\usepackage{textcomp}
\usepackage{stfloats}
\usepackage{url}
\usepackage{verbatim}
\usepackage{graphicx}
\usepackage{epstopdf}
\usepackage{cite}
\usepackage{cuted}
\usepackage{multirow}
\usepackage{dsfont}
\usepackage{bbm}
\usepackage{amsmath}
\newtheorem{remark}{Remark}
\allowdisplaybreaks[4]
\footnoterule

\usepackage{geometry}
\def\BibTeX{{\rm B\kern-.05em{\sc i\kern-.025em b}\kern-.08em
    T\kern-.1667em\lower.7ex\hbox{E}\kern-.125emX}}
\def\BibTeX{{\rm B\kern-.05em{\sc i\kern-.025em b}\kern-.08em
    T\kern-.1667em\lower.7ex\hbox{E}\kern-.125emX}}

\begin{document}

\title{Two-Way Semantic Transmission of Images without Feedback
}%
\author{\IEEEauthorblockN{Kaiwen Yu, Qi He, and Gang Wu}
%
\thanks{This work was supported by the by the National Key R\&D Program of China under Grant 2020YFB1806604.}
\thanks{K. Yu, Q. He and G. Wu are with the National Key Laboratory of Science and Technology on Communications, University of Electronic Science and Technology of China, Chengdu 611731, China(e-mail: yukaiwen@std.uestc.edu.cn, heqi@uestc.edu.cn, wugang99@uestc.edu.cn).}
}

\maketitle

\begin{abstract}
  As a competitive technology for 6G, semantic communications can significantly improve transmission efficiency. However, many existing semantic communication systems require information feedback during the training coding process, resulting in a significant communication overhead. In this article, we consider a two-way semantic communication (TW-SC) system, where information feedback can be omitted by exploiting the weight reciprocity in the transceiver. Particularly, the channel simulator and semantic transceiver are implemented on both TW-SC nodes and the channel distribution is modeled by a conditional generative adversarial network. Simulation results demonstrate that the proposed TW-SC system performs closing to the state-of-the-art one-way semantic communication systems but requiring no feedback between the transceiver in the training process.
\end{abstract}


\section{Introduction}

\IEEEPARstart {W}{ith} the rapid development of industrial internets and internets of vehicles, the demand on content and control information transmission has increased dramatically\cite{ad1}. Different from traditional communications that focus on recovering the transmitted symbols as accurate as possible, semantic communication recovers the semantic meaning of the transmitted content\cite{j1,d1}, which requires much less communication resource and is expected to break through the bottleneck of the existing systems and become a very competitive 6G technology\cite{a2}.
The existing research on semantic communications mainly focuses on the construction of a one-way or unidirectional communication system while the two-way semantic communications are required in many situations. For example, heterogeneous robots collaborate to complete a task or vehicles cooperate with each other for autonomous driving \cite{ex}. Bidirectional intelligent communication tasks inspire us to study two-way semantic communications.

Assisted by deep learning methods, semantic communication systems can perform effectively in text transmission\cite{c1,d2,a3}, speech transmission \cite{c2}, image transmission \cite{c3,i1}, intelligence tasks\cite{j2,j3}, etc.
Particularly, with the joint source-channel coding (JSCC) scheme, the transceiver can be learned in an end-to-end manner to fit the current channel condition, and achieve a better and more robust performance than modular-based systems \cite{c3}. However, in the training process of JSCC, the gradients of neural networks (NNs) should be back propagated from the receiver to the transmitter, which incurs high amount of data feedback. The offline training and online deployment strategy is an attractive option \cite{c1,d2,a3}, while a non-negligible degradation will arise if the channel state used for offline training does not match the current situation\cite{c3,f1}.
Taking image transmission as an example, the key metric, peak signal-to-noise ratio (PSNR), drops about 4 dB compared with online training if the current channel condition is 10 dB while the model trained under 0 dB\cite{f1}. Therefore, many existing works have studied online training to obtain the optimal transceiver or just adapt to the current channel condition to improve performance by fine-tuning\cite{f2,f3,f4}, which however could introduce extra computational and communication overhead\cite{j2}.

Generative adversarial networks (GANs) have been proposed to achieve efficient bit transmission under unknown channel models\cite{f5,c4}. By obtaining an equivalent NN of a random channel using GAN, the optimal parameter weights of the transmitter are learned at the receiver and sent back to the transmitter. Thus, another reliable reversed communication link is required in training.
Simply extending the JSCC in one-way transmission technique to the two-way communication would bring in significant communication and delay overhead. Thus, efficient and practical two-way semantic communication techniques need to be further explored.

This article explores the two-way semantic communication, named TW-SC, where transceiver NNs and channel NN are jointly designed to perform the bidirectional image transmission. Specifically, we use the conditional GAN (CGAN) as the channel simulator to model the wireless channel distribution for image transmission, where the received semantic pilot is treated as additional information to CGAN, referred to as SP-CGAN. By utilizing the \textit{weight reciprocity} between both sides, TW-SC can be learned locally and efficiently without any gradient exchange. It is worthy noting that the TW-SC framework can be easily extended to the bidirectional transmission of other types of data. Based on our simulation results, the proposed TW-SC system performs closing to that of the state-of-the-art JSCC-based one-way image transmission systems without requiring the overhead on data exchange for training. The codes are available at: https://github.com/Kiven-ykw/TW-SemanticComm.

\section{System and Learning Models}
In this section, we introduce the proposed two-way semantic communication system, including communication model, channel modeling, and TW-SC.

\subsection{Communication Model}

\begin{figure}[htbp]
\centerline{\includegraphics[width=3.5in]{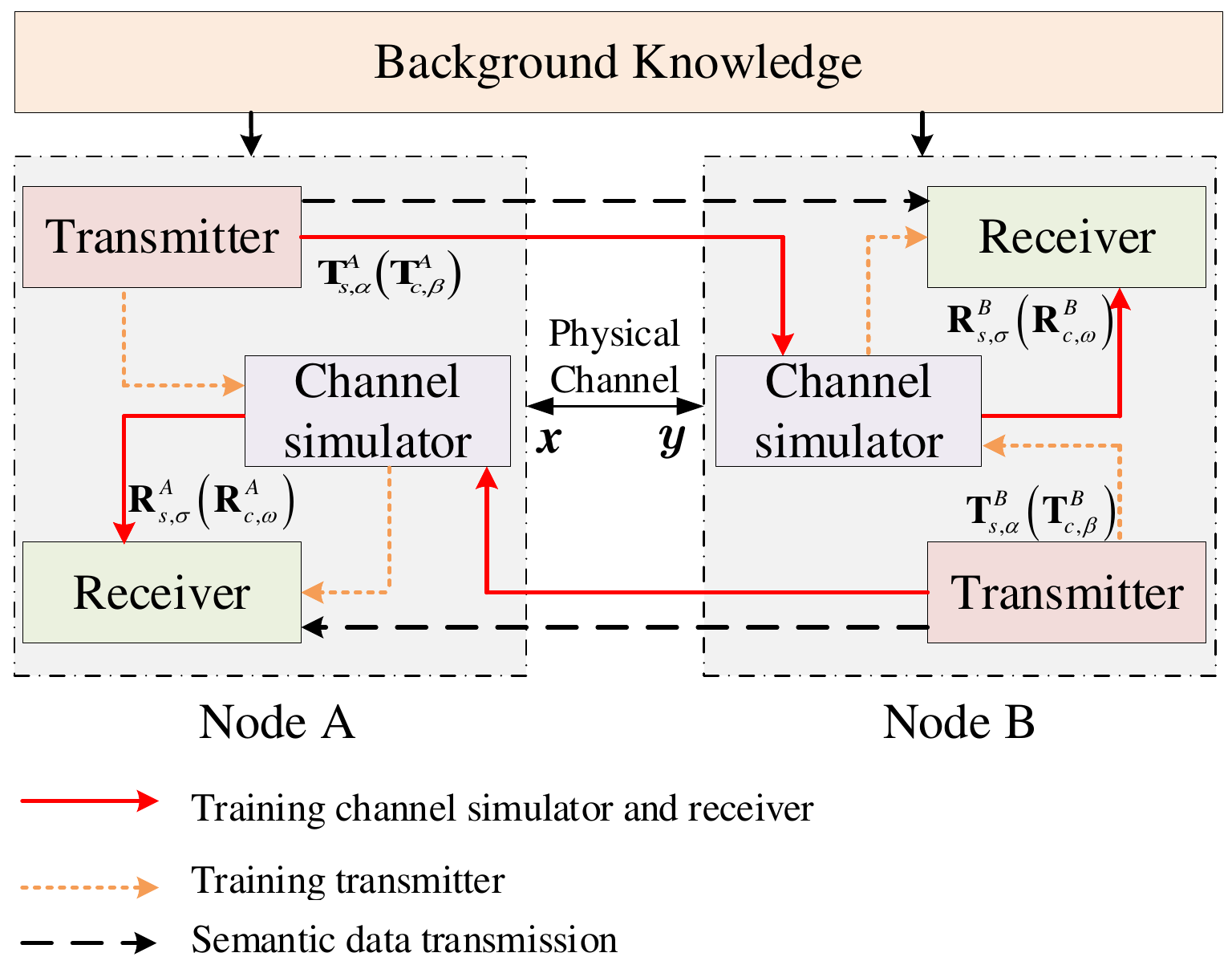}}
\caption{The structure of proposed TW-SC.}
\label{training}
\end{figure}

We consider that nodes A and B exchange image messages $\pmb M^A, \pmb M^B \in \mathbb R^{H \times W \times C}$, where $H$, $W$ and $C$ are the image height, width, and channel, respectively. i.e., node A sends $\pmb M^A$ to B, and node B sends $\pmb M^B$ to A. The channel simulator is used to learn the wireless channel locally. The transmitter consists of semantic encoding and channel encoding, wherein semantic encoding converts the input image, $\pmb M^J, J \in \{A,B\}$, into symbols $\pmb x^J$, that are coded by the channel encoding and transmitted over the physical wireless channel. Take the transmission from node A to node B as an example,
\begin{equation}\label{Tx}
{\pmb x^A} = {\bf{T}}_{c,\beta }^A\left( {{\bf{T}}_{s,\alpha }^A\left( {{\pmb M^A}} \right)} \right),
\end{equation}
where ${\bf{T}}_{c,\beta }^A\left( \cdot\right)$ and ${\bf{T}}_{s,\alpha }^A\left( \cdot\right)$ denote the channel and semantic encoding NN with weight $\beta$ and $\alpha$, respectively. The received signal is represented by
\begin{equation}\label{Rx}
\pmb {y}^B = {\mathbf{H}}{\pmb x^A} + \mathbf{n},
\end{equation}
where $\mathbf{H}$ is the wireless physical channel and $n \in \mathcal N\left( {0,{\sigma ^2}} \right)$ is the additive circular complex Gaussian noise with zero mean and variance $\sigma^2$. We assume that the channels between nodes A and B are completely reciprocal such as in the time division duplexing (TDD) communications, that is the channel from node A to node B is identical to that from B to A, e.g. ${\mathbf{H}}_{B}={\mathbf{H}}_{A}$. Denote the complex channel output as $\pmb{y}$, that is $\pmb y = \mathbf H_{A,B}\pmb x + \mathbf n$. Similarly, the receiver is also composed of semantic decoding and channel decoding, which aims to recover the original image. As a result, the recovered signal can be expressed as,
\begin{equation}\label{Rx}
{\hat {\pmb M}^B} = {\bf{R}}_{s,\sigma }^B\left( {{\bf{R}}_{c,\omega }^B\left( {{\pmb{y}^B}} \right)} \right),
\end{equation}
where ${\bf{R}}_{c,\omega }^B\left( \cdot \right)$ and ${\bf{R}}_{s,\sigma }^B\left( \cdot \right)$ denote the channel and semantic decoding NN with weight $\omega$ and $\sigma$, respectively. Assuming there is no mutual interference between the two links, such as assigned with different time slots, the optimal goal of the designed TW-SC is to minimize the semantic errors in two-way communications, which can be formulated as
\begin{equation}\label{goal}
  \mathop {\min }\limits_{{\bf{T}}_{c,\beta }^J,{\bf{T}}_{s,\alpha }^J,{\bf{R}}_{c,\omega }^J,{\bf{R}}_{s,\sigma }^J} \frac{1}{2}\left( {\underbrace{\left\| {{{\hat{ \pmb M}}^A} - {\pmb M^B}} \right\|_2^F}_{\text{Node B to A}} + \underbrace{\left\| {{{\hat {\pmb M}}^B} - {\pmb M^A}} \right\|_2^F}_{\text{Node A to B}}} \right).
\end{equation}

\begin{algorithm}[t]\label{alg1}
\caption{The designed SP-CGAN training algorithm for channel simulator.}
\begin{algorithmic}[1]

\STATE \textbf{Initialize:} Initialize the weights ${\bf{G}}_{\theta}^J$, ${\bf{D}}_{\eta }^J$, $J \in \{A, B\}$;

\STATE \textbf{Input:} The node A and B transmit images $\pmb M$.

\STATE \textbf{Output:} The whole network ${\bf{G}}_{\theta}^J$, ${\bf{D}}_{\eta }^J$, $J \in \{A, B\}$;

\STATE \textbf{Training Generator:}

\STATE Sample transmit image $\pmb M$ and channel $\bf H$.
\STATE Put the output of transmitter $\pmb x^J$ into condition information.
\STATE Get the output of the received semantic pilot signal from the channel $\bf H$.
\STATE Get the samples noise $\bf z$
\STATE Update the generator by minimizing the loss function (\ref{gen})

\STATE \textbf{Training Discriminator:}
\STATE Sample transmit image $\pmb M$ and channel $\bf H$.
\STATE Put the output of transmitter $\pmb x^J$ into condition information.
\STATE Get the output of the received semantic pilot signal from the channel $\bf H$.
\STATE Get the examples received real data $\pmb y$.
\STATE Update the discriminator by minimizing the loss function (\ref{disc})
\end{algorithmic}
\end{algorithm}

\subsection{Modeling channel distribution}
\begin{figure}[htbp]
\centerline{\includegraphics[width=3.5in]{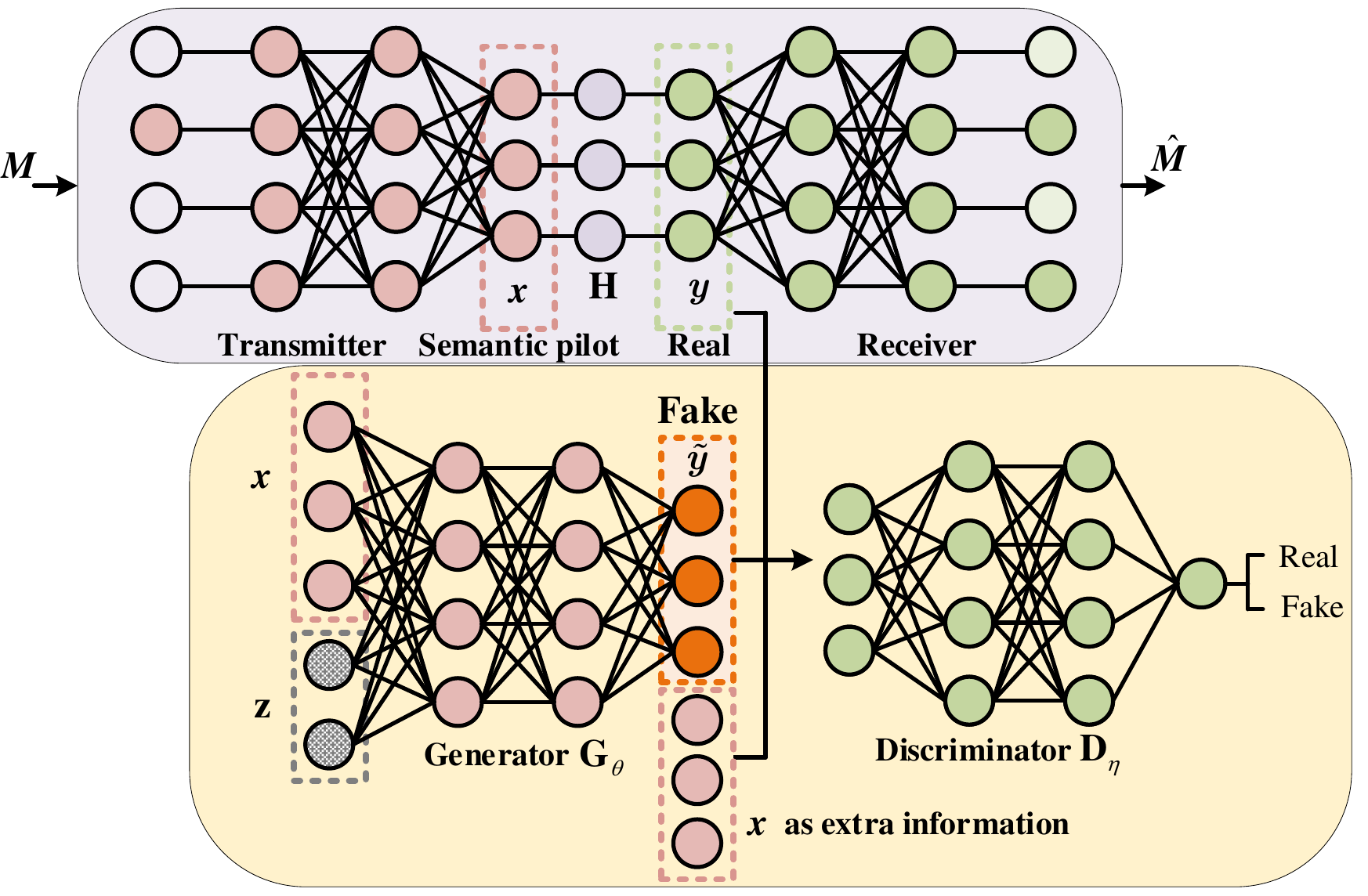}}
\caption{The details of training SP-CGAN.}
\label{trainingDetail}
\end{figure}

Motivated by \cite{c4}, we use CGAN to model the distribution of wireless channel $\mathbf H$.
The received semantic features are used as \textit{semantic pilot} as extra information to the generator and discriminator, dubbed SP-CGAN. The loss functions of the generator and discriminator are depicted as
\begin{equation}\label{gen}
{\mathcal L_g} = \mathop {\min }\limits_{{\mathbf G_\theta}} \mathbb{E}[\max\{ 1 - {{\mathbf D_\eta}}(\mathbf G_\theta({\mathbf z}),0\} ],
\end{equation}
\begin{equation}\label{disc}
{\mathcal L_d} = \mathop {\min }\limits_{{\mathbf D_\eta}} \mathbb{E}[ - {\mathbf D_\eta}({\mathbf H}) - \max\{ 1 - \mathbf D_\eta (\mathbf G_\theta({\mathbf z})),0\} ],
\end{equation}
where $\mathbf G_\theta$ and $\mathbf D_\eta$ denote the generator NN with weight $\theta$ and the discriminator NN with weight $\eta$, respectively. $\mathbf D_\eta(\mathbf H)$ and $\mathbf D_\eta (\mathbf G_\theta(\mathbf z))$ denote the probabilities that the discriminator considers that the sample is from a real channel or generated by a NN, respectively. $\bf z$ denotes the sample noise. The $\mathbf G_\theta$ and $\mathbf D_\eta$ can be optimized by modeling as a two-player maximum minimization game problem, which is depicted as
\begin{equation}\label{obj}
   \mathop {\min }\limits_{{\mathbf G_\theta }} \mathop {\max }\limits_{{\mathbf D_\eta }} \mathbb E\left[ {{\mathbf D_\eta }\left( \mathbf H \right)} \right] + \mathbb E\left[ {\max \left\{ {1 - {\mathbf D_\eta }\left( {{\mathbf G_\theta }\left( \mathbf z \right)} \right),0} \right\}} \right].
\end{equation}
The training algorithm for modeling the wireless channel distribution with the designed SP-CGAN is shown in Fig.~\ref{trainingDetail} and Algorithm~1, where the NNs of the generator and the discriminator are trained iteratively in each iteration, that is, the parameters of one model will be fixed while training the other model. With the learned transmitter and receiver, the encoded semantic information from the transmitter through the actual channel can be used to obtain the true data whilst the encoded semantic information through the channel generator is used to obtain the fake data in the receiver. According to the loss functions of equations (\ref{gen}) and (\ref{disc}), the parameters of the generator and discriminator are updated, respectively.

\subsection{TW-SC training without information feedback}

\begin{algorithm}[t]
\caption{TW-SC Training Algorithm.}
\begin{algorithmic}[1]

\STATE \textbf{Initialize:} Initialize the weights ${\bf{T}}_{c,\beta }^J\left( \right)$, ${\bf{T}}_{s,\alpha }^J\left( \right)$, ${\bf{R}}_{c,\omega }^J\left( {} \right)$, ${\bf{R}}_{s,\sigma }^J$, $J \in \{A, B\}$;

\STATE \textbf{Input:} The node A and B transmit images $\pmb M$.

\STATE \textbf{Output:} The whole network ${\bf{T}}_{c,\beta }^J\left( \right)$, ${\bf{T}}_{s,\alpha }^J\left( \right)$, ${\bf{R}}_{c,\omega }^J\left( {} \right)$, ${\bf{R}}_{s,\sigma }^J$, $J \in \{A, B\}$;

\STATE \textbf{Training SP-CGAN and receivers cross nodes:}

\STATE Semantic encoding: ${\bf{T}}_{s,\alpha }^J\left( \pmb M \right)$ $\rightarrow$ $\pmb s$,
\STATE Channel encoding: ${\bf{T}}_{c,\beta }^J\left( \pmb s \right)$ $\rightarrow$ $\pmb x$,
\STATE Though the real channel: $\pmb y^J = {\mathbf{H}}{\pmb x^J} + \pmb{n}$,
\STATE Channel decoding: ${\bf{R}}_{c,\omega }^J\left( {\pmb y^J} \right)$ $\rightarrow$ $\pmb {\hat{s}}$
\STATE Semantic decoding: ${\bf{R}}_{s,\sigma }^J\left( \pmb {\hat{s}} \right)$ $\rightarrow$ $\pmb {\hat{M}}$
\STATE Update the SP-CGAN by \textbf{Algorithm~1}
\STATE Update the receivers by minimizing the loss function (\ref{lossA}) and (\ref{lossB})
\STATE \textbf{Training transmitters at local node:}

\STATE Semantic encoding: ${\bf{T}}_{s,\alpha }^J\left( \pmb M \right)$ $\rightarrow$ $\pmb s$,
\STATE Channel encoding: ${\bf{T}}_{c,\beta }^J\left( \pmb s \right)$ $\rightarrow$ $\pmb x$,
\STATE Though the channel simulator \textit{SP-CGAN}: ${\bf{G}}_{\theta}^J\left( \pmb x \right)$ $\rightarrow$ $\pmb y$;
\STATE Channel decoding: ${\bf{R}}_{c,\omega }^J\left( {\pmb y} \right)$ $\rightarrow$ $\pmb {\hat{s}}$
\STATE Semantic decoding: ${\bf{R}}_{s,\sigma }^J\left( \pmb {\hat{s}} \right)$ $\rightarrow$ $\pmb {\hat{M}}$
\STATE Update the transmitters by minimizing the loss function (\ref{lossTA}) and (\ref{lossTB})

\label{code:algorithm2}
\end{algorithmic}
\end{algorithm}

The proposed TW-SC algorithm is conducted in two stages, as illustrated in Fig. \ref{training}. In the first stage, represented by the solid red lines, transmitters at nodes A and B send training data to train the channel simulator and receiver to ensure convergence. In the second stage, represented by the dotted orange lines, the transmitters send training data inside the local node to train the transmitter, avoiding gradient back propagation over the wireless channel. The training data known at nodes A and B is represented by $\pmb M$. In the NN training, we adopt mean-squared error (MSE) as the end-to-end loss function, which can be expressed as,
\begin{equation}\label{lossA}
 \mathcal L_R^A(\pmb M^B,\hat { \pmb M}^A) = \frac{1}{{hw}}\sum\limits_{i = 1}^h {\sum\limits_{j = 1}^w {{{\left( {{\pmb M_{i,j}^B} - {{\hat { \pmb M}}_{i,j}^A}} \right)}^2}} },
\end{equation}
\begin{equation}\label{lossB}
 \mathcal L_R^B(\pmb M^A,\hat { \pmb M}^B) = \frac{1}{{hw}}\sum\limits_{i = 1}^h {\sum\limits_{j = 1}^w {{{\left( {{\pmb M_{i,j}^A} - {{\hat { \pmb M}}_{i,j}^B}} \right)}^2}} },
\end{equation}
respectively, where $\pmb M_{i,j}^J, j \in \{A, B\}$ denotes the pixel value at position $(i,j)$ of the image, $h \times w$ is the image size. Note that the receiver's gradient calculation is performed locally by the gradient descent approach. Without the back propagation of gradients, the transmitter cannot obtain the loss information. We investigate the characteristics of TW-SC in order to get answers to this issue.
\begin{remark}
\textit{\textbf{Weight Reciprocity:} When the channels are reciprocal and the NNs of node A and node B have the same hyper parameters, the receivers of node A and node B have the same gradient, which helps us to complete the training of transmitters locally and does not require inter-node feedback, thus effectively reducing the communication overhead and delay.}
\end{remark}


The gradient of the transmitter can be determined locally using the \textit{Weight Reciprocity} and SP-CGAN. The loss function at the transmitter can be written as
\begin{equation}\label{lossTA}
\mathcal L_T^A = \mathbb E\left\{ {l\left( {\pmb M,{\bf{R}}_{s,\sigma }^A\left( {{\bf{R}}_{c,\omega }^A\left( {{{\bf {G}}_\theta ^A}\left( {{\bf C}^B,\bf z} \right)} \right)} \right)} \right)} \right\},
\end{equation}
\begin{equation}\label{lossTB}
\mathcal L_T^B = \mathbb E\left\{ {l\left( {\pmb M,{\bf{R}}_{s,\sigma }^B\left( {{\bf{R}}_{c,\omega }^B\left( {{{\bf {G}}_\theta ^B}\left( {{\bf C}^A,\bf z} \right)} \right)} \right)} \right)} \right\},
\end{equation}
where $l$ denotes the MSE loss function, ${\bf C}^J = {\bf{T}}_{c,\beta }^J\left( {{\bf{T}}_{s,\alpha }^J\left( {{\pmb M^J}} \right)} \right), J \in \{A,B\}$.
Transmitters A and B are able to train the weights via their own internal receivers by minimizing the loss functions, (\ref{lossTA}) and (\ref{lossTB}), inside nodes A and B. The gradient loss can be internally back propagated to achieve convergence using the channel simulator SP-CGAN. Thus, the proposed method completely avoids random wireless channel, saves communication overhead, and reduces training latency, and is applicable to both online real-time and offline training. The proposed TW-SC algorithm is summarized in Algorithm 2.

\section{Numerical Results}
In this part, we evaluate the proposed TW-SC in terms of transmission performance. We use the MNIST handwritten digit dataset for the experiments, where the training set consists 60,000 elements and the test set includes 10,000 elements. In the experiment, the semantic encoding consists of five two-dimensional convolutional layers (Conv2D) with 4, 8, 8, 16, and 16 filters. The channel encoding consists of four Conv2D with 16, 16, 32, and 32 filters. The Conv2Ds are with a $3 \times 3$ kernels and followed by an ELU activation function. The channel decoding consists of four Transpose Conv2Ds (TransConv2D) with 32, 32, 16, and 16 filters and the semantic decoding consists of five TransConv2Ds with 8, 8, 4, 4, and 1 filters. The TransConv2Ds are also with a $3 \times 3$ kernels and followed by an ELU activation function. For the SP-CGAN, the generator consists of four Conv1Ds with 256, 128, 64, and 2 filters. The four Conv1Ds are with $5 \times 5$, $3 \times 3$, $3 \times 3$, and $3 \times 3$ kernels. Similarly, the discriminator consists of four Conv1Ds with 256, 128, 64, and 16 filters and with $5 \times 5$, $3 \times 3$, $3 \times 3$, and $3 \times 3$ kernels and a dense layer with 100 filters. The ReLU function is selected as the activation function. For all networks, the learning rate is $\delta=1e^{-3}$ and learning rate decay factor is ${\delta_d}=1e^{-4}$. The weights of models are updated by the Adam optimizer and the epoch for training is 100. Moreover, we set the training batch size to 128 and compare the proposed scheme with the two semantic communication algorithms under the AWGN channel and Rayleigh fading channel, respectively,
\begin{itemize}
  \item One-way end-to-end JSCC: It is trained under real channels and is used as the optimal algorithm.
  \item One-way end-to-end semantic communication with unknown channels: This method simulates the channel using GAN and then conducts transceiver training according to the simulated channel, dubbed GAN-SC. Semantic pilot is not used by GAN as extra data.
\end{itemize}
To make a fair comparison with the two benchmarks, we average the link performance of the proposed TW-SC scheme. In addition, the widely recognized evaluation metrics in computer vision, namely structure similarity (SSIM) and PSNR, are used to measure the performance of the three algorithms. The suffixes `-AWGN' and `-Rayleigh' indicate cases trained under the AWGN and Rayleigh channels, respectively.

\begin{figure}[htbp]
\centerline{\includegraphics[width=2.3in]{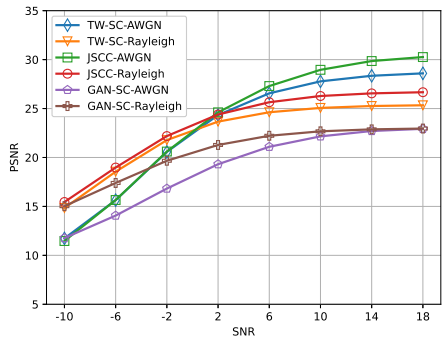}}
\caption{The PSNR score versus SNR in AWGN channel.}
\label{PSNR-TW-SC-AWGN}
\end{figure}

Fig.~\ref{PSNR-TW-SC-AWGN} shows the relationship between the PSNR score and SNR over the AWGN channel. For the AWGN channel, the suggested TW-SC-AWGN performs well and gets close to the optimal algorithm at the low SNR range, which verifies the efficiency of \textit{weight reciprocity} in the two-way system. The GAN-SC, based on the GAN simulated channel, performs poorest. We also examine the performance of methods in comparison developed under the Rayleigh channel to confirm the algorithm's generalization ability. The TW-SC trained under the AWGN channels outperforms the model trained under the Rayleigh channels. From the figure, the model trained under a different channel from the test channel suffers from some performance degradation.
\begin{figure}[htbp]
\centerline{\includegraphics[width=2.3in]{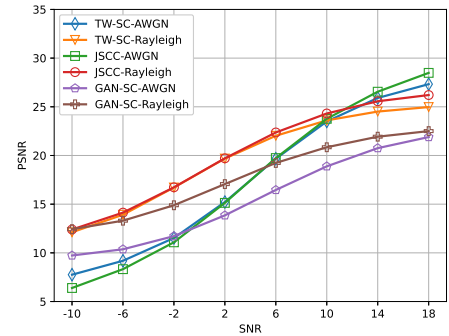}}
\caption{The PSNR score versus SNR in Rayleigh channel.}
\label{PSNR-TW-SC-Ray}
\end{figure}

Fig.~\ref{PSNR-TW-SC-Ray} shows the relationship between the PSNR score and SNR over the Rayleigh fading channel. Obviously, Rayleigh fading channel has more negative effect than the AWGN channel for the three methods. When SNR is lower than 10 dB, the proposed TW-SC trained under the Rayleigh channels has a higher PSNR performance than the models trained in the AWGN channels. Besides, the proposed TW-SC trained under the AWGN channels has a wonderful performance close to the optimal JSCC, which proves the validity of the proposed algorithm. On the other hand, the GAN-SC algorithm also has the worst performance since the GAN generates random data that obeys that distribution. In contrast, SP-CGAN is able to generate channel-specific distributions with the aid of semantic pilots, which demonstrates the necessity of pilot as additional information and the effectiveness of the proposed SP-CGAN algorithm.
\begin{figure}[htbp]
\centerline{\includegraphics[width=2.3in]{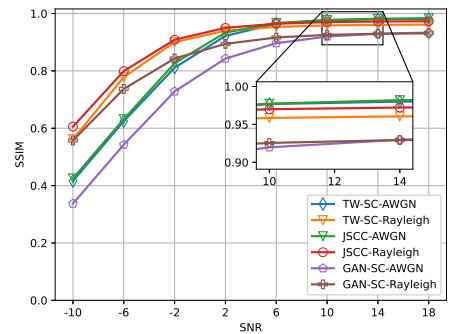}}
\caption{The SSIM score versus SNR in AWGN channel.}
\label{SSIM-TW-SC-AWGN}
\end{figure}

Fig.~\ref{SSIM-TW-SC-AWGN} demonstrates the relationship between the SNR and SSIM scores over the AWGN channel. From the figure, when SNR is lower than 6 dB, the models trained under the Rayleigh fading channels get the high SSIM performance, which is in line with existing studies that training under the fading channel can improve the robustness of models over various channel types \cite{c2}. The JSCC-AWGN schemes yield higher SSIM score than other schemes, when SNR is higher than 6 dB. Given the semantic pilot as additional information for CGAN, the proposed TW-SC performs steadily when coping with different fading channels and SNRs. However, for the GAN-SC, the performance is quite poor under dynamic channel conditions, especially in the low SNR regime.
\begin{figure}[htbp]
\centerline{\includegraphics[width=2.3in]{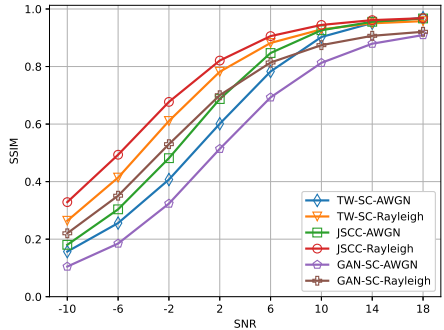}}
\caption{The SSIM score versus SNR in Rayleigh channel.}
\label{SSIM-TW-SC-Ray}
\end{figure}
Fig.~\ref{SSIM-TW-SC-Ray} compares the SSIM performance between TW-SC and the two benchmarks over the Rayleigh channel. The JSCC-Rayleigh gets the optimal performance. The proposed TW-SC achieves a very high SSIM score that is close to the optimal JSCC algorithm. As the SNR increases, the performance of the proposed algorithm almost overlaps with the optimal algorithm.

\begin{figure}[htbp]
\centerline{\includegraphics[width=2.3in]{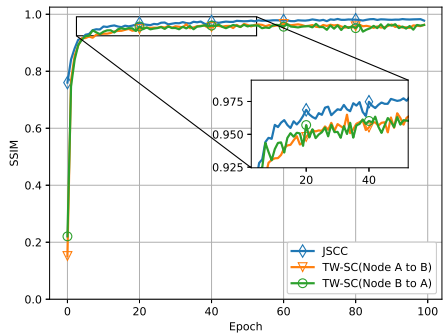}}
\caption{The SSIM score versus Epoch in AWGN channel.}
\label{SSIMvsEpoch}
\end{figure}
Fig.~\ref{SSIMvsEpoch} demonstrates the relationship between the training epoch and the SSIM score over the AWGN channel, which enables more intuitive verification of the reliability of the weight reciprocity and the high convergence speed of TW-SC. From the figure, all the three curves converge within 20 epochs. Meanwhile, the convergence speeds of links from node A to node B and from B to A are very similar and approximate to the ideal JSCC algorithm, which shows that the proposed training scheme can have the ability to train the transmitter without feedback while obtaining a performance close to the optimal algorithm.

\section{Conclusion}
In this article, we have studied the two-way semantic communication system and proposed a training scheme for the semantic transceiver locally by learning the channel distribution, thus enabling the training of the transceiver without information feedback and model transfer. The simulation results show that our proposed two-way semantic communication scheme performs close to the bidirectional scheme consisting of two independent optimal one-way semantic communication systems but gets rid of the feedback overhead in training.

\vspace{12pt}
\vfill

\end{document}